\documentclass[twocolumn, amssymb, amsmath, aps, superscriptaddress, showpacs, footinbib, prb]{revtex4}
\usepackage{graphics}

\newcommand{\compound}{Pb$_{0.55}$Cd$_{0.45}$V$_2$O$_5$}
\newcommand{\afm}{\text{AFM}}
\newcommand{\fm}{\text{FM}}

\begin{document}

\title{Spin ladder compound Pb$_{0.55}$Cd$_{0.45}$V$_2$O$_5$: synthesis and investigation}
\date{\today}
\author{Alexander~A.~Tsirlin}
\email{tsirlin@icr.chem.msu.ru}
\affiliation{Department of Chemistry, Moscow State University, 119992, Moscow, Russia}
\affiliation{Max-Planck Institute CPfS, N\"othnitzer Str.~40, 01187, Dresden, Germany}
\author{Roman~V.~Shpanchenko}
\email{shpanchenko@icr.chem.msu.ru}
\author{Evgeny~V.~Antipov}
\affiliation{Department of Chemistry, Moscow State University, 119992, Moscow, Russia}
\author{Catherine~Bougerol}
\affiliation{Equipe CEA-CNRS NPSC SP2M, DRFMC, CEA, 17 rue des Martyrs, 38054 Grenoble, France}
\author{Joke~Hadermann}
\author{Gustaaf~\surname{Van Tendeloo}}
\affiliation{EMAT University of Antwerp (RUCA), Groenenborgerlaan 171, 2020, Antwerp, Belgium}
\author{Walter~Schnelle}
\author{Helge~Rosner}
\email{rosner@cpfs.mpg.de}
\affiliation{Max-Planck Institute CPfS, N\"othnitzer Str.~40, 01187, Dresden, Germany}

\begin{abstract}
The complex oxide Pb$_{0.55}$Cd$_{0.45}$V$_2$O$_5$ was synthesized and investigated by means of X-ray powder diffraction, electron diffraction, magnetic susceptibility measurements and band structure calculations. Its structure is similar to that of MV$_2$O$_5$ compounds (M = Na, Ca) giving rise to a spin system of coupled $S=1/2$ two-leg ladders. Magnetic susceptibility measurements reveal a spin gap-like behavior with $\Delta\approx 270$ K and a spin singlet ground state. Band structure calculations suggest Pb$_{0.55}$Cd$_{0.45}$V$_2$O$_5$ to be a system of weakly coupled dimers in perfect agreement with the experimental data. Pb$_{0.55}$Cd$_{0.45}$V$_2$O$_5$ provides an example of the modification of the spin system in layered vanadium oxides by cation substitution. Simple correlations between the cation size, geometrical parameters and exchange integrals for the MV$_2$O$_5$-type oxides are established and discussed.
\end{abstract}

\maketitle

\section{Introduction}
Complex vanadium oxides often present attractive examples of low-dimensional magnetic systems and exhibit unusual phenomena like spin gap formation or a spin liquid ground state.\cite{ueda,vasiliev} These phenomena are not fully understood yet, therefore the search and investigation of novel compounds revealing low-dimensional spin systems is still a challenging task for solid state physics. Basically, there are two possible ways to search for new spin systems. The first way deals with the study of new structural types that are completely different from those of well-known and thoroughly investigated compounds. The second way suggests a systematic investigation of a number of related compounds. Both ways have been successfully realized during the last decade.\cite{na2v3o7,li2vosio4,ueda} However, the search for new structural types requires substantial intuition and luck. Systematic studies are more predictable if one can understand the relationship between the composition, crystal structure of the solid and the magnetic interactions in it.

\begin{figure}[hb!]
\includegraphics{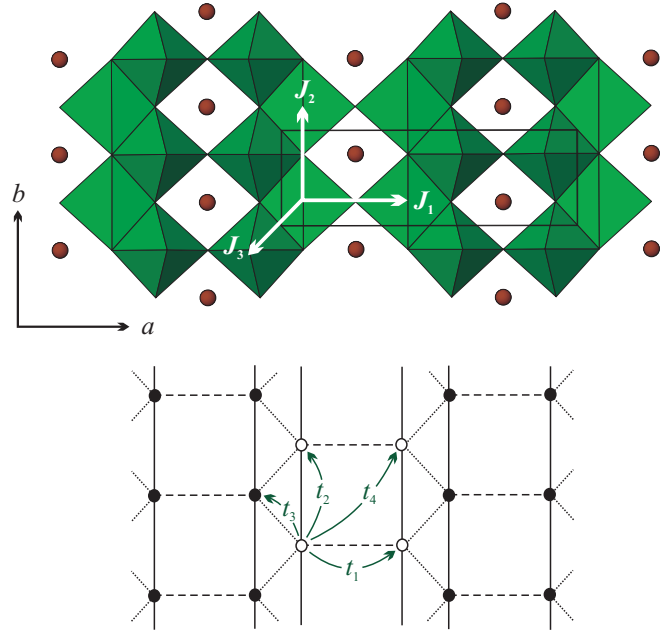}
\caption{\label{structure}(Color online) The projection of the \compound\ crystal structure along the $c$ axis (upper panel) and the system of coupled two-leg spin ladders (lower panel). In the upper panel circles denote the mixed Pb/Cd position, white arrows show three different nearest-neighbor exchange interactions. In the lower panel solid lines correspond to the legs of the ladders, dashed lines show the rungs and dots indicate inter-ladder coupling. The inter-layer hopping $t_{\perp}$ is not shown.}
\end{figure}
MV$_2$O$_5$ compounds (M is an alkaline or alkaline-earth metal) are promising candidates for a systematic study of magnetic interactions in vanadium oxides. All of them (except for that with M = Cs) have layered structures with layers formed by edge- and corner-sharing VO$_5$ square pyramids (see Fig.~\ref{structure}). The M cations are situated between the layers. If M is a monovalent cation the average oxidation state of vanadium is +4.5 and both charge ordered (LiV$_2$O$_5$) or disordered (NaV$_2$O$_5$ above 35 K) states may be realized along with 1D magnetic behavior.\cite{ueda,liv2o5_elstructure,liv2o5,charge,liv2o5_nmr} Actually, the properties of NaV$_2$O$_5$ are even more complicated since different patterns of charge ordering have been recently found for this compound at low temperature.\cite{nav2o5_order,nav2o5_devil} 

The physics of MV$_2$O$_5$ compounds with divalent M cations (Ca, Mg) is simpler due to the absence of charge degrees of freedom. In both oxides vanadium atoms are tetravalent and a system of coupled $S=1/2$ two-leg spin ladders is realized. However, exchange interactions in CaV$_2$O$_5$ and MgV$_2$O$_5$ are quite different. In CaV$_2$O$_5$ the strongest interaction runs along the rung of the ladder ($J_1\approx 600$ K, see Fig.~\ref{structure}) while other interactions ($J_2,\ J_3,\ J_4$) do not exceed 100 -- 150 K.\cite{korotin,korotin_prl,cav2o5_qmc,cav2o5_nmr} Thus, weakly coupled spin dimers are formed and CaV$_2$O$_5$ reveals a spin gap $\Delta\approx J_1$ [experimental values of $\Delta$ range from 560 to 660 K (Refs. \onlinecite{onoda,raman,cav2o5_nmr})]. In contrast, MgV$_2$O$_5$ reveals comparable nearest-neighbor interactions $J_1,\ J_2,\ J_3$ of about 100 K.\cite{korotin,korotin_prl} The presence of a spin gap in MgV$_2$O$_5$ is still argued and the suggested values of $\Delta$ are quite small ($15-20$ K).\cite{mgv2o5_prb,mgv2o5_japan,onoda-mg,raman} The change of V--O--V angles has been claimed in Ref.~\onlinecite{korotin} as the main reason for the differences between CaV$_2$O$_5$ and MgV$_2$O$_5$.

A general phase diagram for the coupled $S=1/2$ two-leg spin ladders is rather complicated revealing unusual ground states (namely, spiral order) and a number of quantum critical points.\cite{diagram} However, most of the regions of this diagram still lack experimental study. The remarkable difference between the magnetic properties of CaV$_2$O$_5$ and MgV$_2$O$_5$ suggests that an appropriate substitution in the M cation position of MV$_2$O$_5$ enables to modify the spin system and to shift between different regions of the phase diagram. Here we present the results of synthesis and investigation of \compound\ -- an example of MV$_2$O$_5$-type compounds combining two different cations in the M position. The joint experimental and computational study of \compound\ allows us to carry out a comparison between several MV$_2$O$_5$-type compounds and to establish correlations between the cation size, geometrical parameters and exchange integrals.

\section{Methods}
A powder sample of \compound\ was obtained by heating a mixture of Pb$_2$V$_2$O$_7$, CdO, V$_2$O$_3$ and V$_2$O$_5$ (ratio 2:4:1:1) in an evacuated and sealed silica tube at 700$^0$C for 24 hours. A change in the Pb:Cd ratio always resulted in the appearence of trace amounts of impurity phases in the samples. A change of heating conditions also led to the formation of impurities.

X-ray powder diffraction (XPD) data for the structure refinement were collected on a STOE diffractometer (transmission mode, CuK$\alpha_1$-radiation, Ge-monochromator, linear-PSD). The JANA2000 program \cite{jana} was used for Rietveld refinement.

Transmission electron microscopy (electron diffraction (ED) and EDXS) was performed using a Philips CM20 microscope with an Oxford Instruments Inca EDXS analyser.

Magnetic susceptibility measurements were carried out using a Quantum Design MPMS SQUID magnetometer in the range between 2 and 400 K at fields $\mu_0H$ of 0.01, 0.2 and 1 T.

Scalar relativistic band-structure calculations were performed using the full-potential non-orthogonal local-orbital minimum-basis scheme\cite{fplo} and the parametrization of Perdew and Wang for the exchange and correlation potential.\cite{perdew} V$(3s,3p,3d,4s,4p)$, Pb$(5s,5p,5d,6s,6p,6d)$, Cd$(4s,4p,4d,5s,5p)$ and O$(2s,2p,3d)$ states, respectively, were chosen as the basis set. All lower-lying states were treated as core states. The inclusion of V$(3s,3p)$, Pb$(5s,5p)$ and Cd$(4s,4p)$ states in the basis set was necessary to account for non-negligible core-core overlaps due to the relatively large extension of the corresponding wave functions. The O $3d$ and Pb $6d$ states were taken into account to get a more complete basis set. A $k$ mesh of 768 points in the Brillouin zone (189 in the irreducible part) was used. The spatial extension of the basis orbitals, controlled by a confining potential\cite{eschrig} $(r/r_0)^4$, was optimized to minimize the total energy. Convergence with respect to the $k$ mesh was carefully checked. A four-band tight-binding (TB) model was fitted to the resulting band structure to find the values of the relevant hopping parameters. The latter values were used to estimate the exchange integrals.

\section{Results}
\subsection{Crystal structure}
\compound\ has an orthorhombic unit cell with lattice parameters $a=11.3565(2)$ \r A, $b=3.6672(1)$ \r A, $c=4.9017(1)$ \r A. Single crystals of \compound\ can not be obtained since above 700$^0$C the compound gradually decomposes without melting. Therefore, the XPD data were used for the structural study.

The lattice parameters for \compound\ and CaV$_2$O$_5$ ($a=11.351$ \r A, $b=3.604$ \r A, $c=4.893$ \r A, space group $Pmmn$ \cite{onoda}) are rather close, therefore we used the crystal structure of CaV$_2$O$_5$ as an initial model for the Rietveld refinement. Indexing of XPD and ED (see below) patterns revealed only the reflection conditions $hk0,\ h+k=2n$ indicating two possible space groups: $Pmmn$ and $P2_1mn$, a subgroup of $Pmmn$. The lead and cadmium atoms were placed into $2b$ position (corresponding to calcium in CaV$_2$O$_5$) and occupancy factors $g_{\text{Pb}}=g_{\text{Cd}}=0.5$ were set. 

\begin{table}
\caption{\label{coordinates}Atomic coordinates and isotropic thermal displacement parameters for \compound}
\begin{ruledtabular}
\begin{tabular}{ccccccc}
Atom & Site & Occupancy & $x$ & $y$ & $z$ & $U_{iso}$ (\r A$^2$) \\
Pb & $2b$ & 0.55(1) & 0.75 & 0.25 & 0.1557(1) & 0.0122(4) \\
Cd & $2b$ & 0.45(1) & 0.75 & 0.25 & 0.1557(1) & 0.0122(4) \\
V & $4f$ & 1 & 0.4063(1) & 0.25 & 0.3883(4) & 0.0076(7) \\
O(1) & $4f$ & 1 & 0.3884(3) & 0.25 & 0.062(1) & 0.008(1) \\
O(2) & $4f$ & 1 & 0.5733(4) & 0.25 & 0.4732(8) & 0.008(1) \\
O(3) & $2a$ & 1 & 0.25 & 0.25 & 0.549(1) & 0.008(1) \\
\end{tabular}
\end{ruledtabular}
\end{table}
The refinement in both $Pmmn$ and $P2_1mn$ space groups gave similar results, therefore the highest possible symmetry ($Pmmn$) was chosen for the final refinement. The atomic displacement parameters of all oxygen atoms were constrained. Anisotropic strain broadening for the profile function was refined to achieve a proper fit for all reflections in the XPD pattern.\cite{foot1} Finally, occupancy factors of lead and cadmium atoms were also refined with a constraint $g_{\text{Pb}}+g_{\text{Cd}}=1$ yielding $g_{\text{Pb}}=0.55(1),\ g_{\text{Cd}}=0.45(1)$ and the \compound\ composition. This composition is in a good agreement with the EDXS result Pb:Cd:V = 1.2(1):1.0(1):3.8(1) but slightly differs from the initial Pb$_{0.5}$Cd$_{0.5}$V$_2$O$_5$ composition. The clear halo in the XPD pattern at $2\theta=15-25^0$ indicates the presence of an amorphous component in the prepared sample. One may roughly estimate the amount of this component by comparing integrated intensities for amorphous (halo) and crystalline (sharp peaks) phases and assuming their scattering powers to be similar. Thus, we find $\sim 25$\% of the amorphous component.

\begin{figure}
\includegraphics{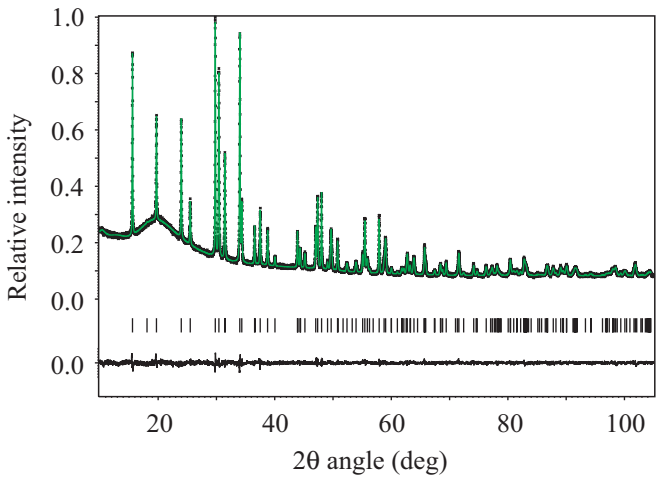}
\caption{\label{rietveld}(Color online) Experimental, calculated and difference XRD patterns for \compound. The halo at $2\theta=15-25^0$ indicates the presence of an amorphous impurity in the sample under investigation.}
\end{figure}

The crystal structure of \compound\ is shown in Fig.~\ref{structure}. The experimental, calculated and difference Rietveld plots are shown in Fig.~\ref{rietveld}. Table \ref{coordinates} presents atomic coordinates and thermal displacement parameters, while Table \ref{distances} summarizes the main interatomic distances and angles in the crystal structure. The final residuals of the refinement are $R_P=0.022,\ R_F=0.020$, and $\chi^2=1.29$.

The crystal structure of \compound\ is very similar to that of CaV$_2$O$_5$. Vanadium atoms form VO$_5$ square pyramids with one short vanadyl bond and four longer equatorial bonds typical for tetravalent vanadium. The V$^{+4}$O$_5$ square pyramids with opposite orientation of vanadyl bonds are linked into zigzag chains via common edges. The pyramids in neighboring chains share their corners and form [V$_2$O$_5$] layers. Lead and cadmium atoms randomly occupy one crystallographic position between the vanadium-oxygen layers. The M--O distances are rather short for lead due to the presence of the smaller cadmium cation in the same position. 

\begin{figure}
\includegraphics{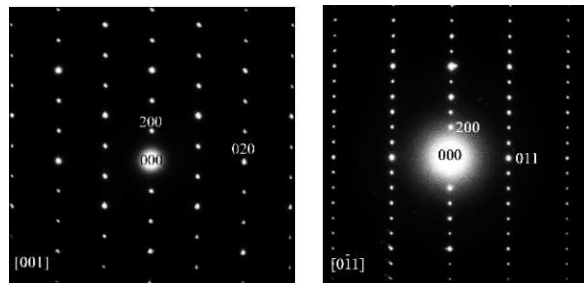}
\caption{\label{ed}Electron diffraction patterns along the [001] and $[0\bar 11]$ directions.}
\end{figure}

\begin{table}
\caption{\label{distances}Selected interatomic distances (in \r A) and angles (in deg). M denotes the mixed Pb/Cd position}
\begin{ruledtabular}
\begin{tabular}{cccc}
M--O(1) & $4\times 2.639(3)$ & V--O(1) & 1.614(5) \\
M--O(2) & $2\times 2.540(4)$ & V--O(2) & 1.942(5) \\
M--O(3) & $2\times 2.336(4)$ & V--O(2) & $2\times 1.969(2)$ \\
& & V--O(3) & 1.942(3) \\
\hline
V--O(2)--V $(J_2)$ & 137.3(2) & V--O(2)--V & 100.88(14) \\
V--O(3)--V $(J_1)$ & 132.1(3) & & \\
\end{tabular}
\end{ruledtabular}
\end{table}
Lead and cadmium cations are fairly different in size (ionic radii 1.45 \r A and 1.21 \r A, respectively), therefore one would expect cation ordering in \compound\ resulting in a decrease of symmetry and superstructure formation. However, all ED patterns (Fig.~\ref{ed}) could be indexed in the $Pmmn$ space group with the cell parameters found from the XPD data. The appearance of the forbidden reflections $h00:h\neq 2n$ in the $[0\bar 11]$ zone is due to double diffraction since these reflections disappear after rotating the crystal away from the perfect orientation. Thus, one may conclude that lead and cadmium atoms in \compound\ are randomly distributed.

We also prepared Pb$_{1-x}$Cd$_x$V$_2$O$_5$ samples with \mbox{$0\leq x\leq 1$}. The MV$_2$O$_5$-type phase was detected for $0.3\leq x\leq 0.7$, but single phase samples were obtained only for $x=0.5$. The lattice parameters for the CaV$_2$O$_5$-type phase were almost identical to those refined for \compound. Thus, the \compound\ phase has no considerable homogeneity range.

\subsection{Magnetic susceptibility}
\begin{figure}
\includegraphics{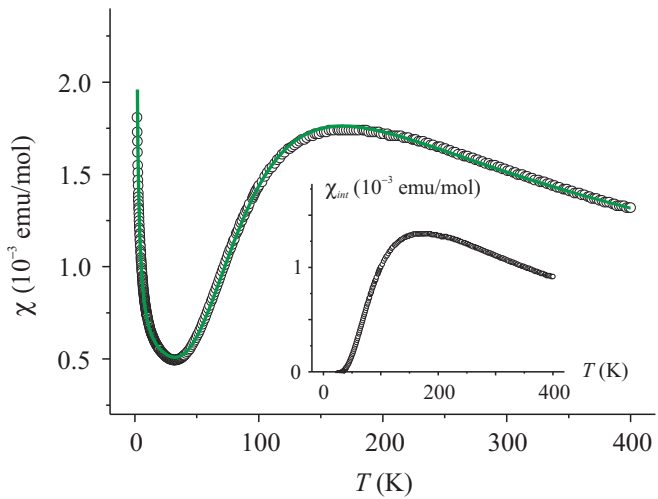}
\caption{\label{suscept}(Color online) Magnetic susceptibility ($\chi_{\text{exp}}$) \textit{vs.} temperature for \compound\ measured at $\mu_0H=0.2$~T. Circles show experimental points while the solid line is the fit with the isolated dimer model. The inset shows the intrinsic susceptibility $\chi_{int}$ of \compound: $\chi_{int}=\chi_{exp}-\chi_0-C/T$ (see equation \ref{eq}).}
\end{figure}

The $\chi(T)$ curve of \compound\ (Fig. \ref{suscept}) reveals a broad maximum, typical for low-dimensional spin systems. The fast decrease of the susceptibility below the maximum may be indicative of a spin gap while the upturn below 30 K is usually attributed to the paramagnetic impact of impurities and defects. In general, the $\chi(T)$ curve of \compound\ is similar to that for CaV$_2$O$_5$.\cite{onoda} Nevertheless, in CaV$_2$O$_5$ the susceptibility maximum is at $T_{\max}\approx 350$ K while in the case of \compound\ $T_{\max}$ is half this value (about 170~K). The position of the susceptibility maximum is usually a characteristic of the strongest exchange interaction in the system. Therefore one may suppose that the spin systems of Pb$_{0.55}$Cd$_{0.45}$V$_2$O$_5$ and CaV$_2$O$_5$ are qualitatively similar, but a slight change of the crystal structure results in a weakening of the magnetic interactions.

CaV$_2$O$_5$ is considered to be a system of weakly coupled dimers, therefore we follow Ref.~\onlinecite{onoda} and fit the experimental curve in Fig. \ref{suscept} with the expression 
\begin{equation}
  \chi=\chi_0+C/T+\dfrac{Ng^2\mu_B^2}{k_BT}\dfrac{1}{e^{\Delta/k_BT}+3}
\label{eq}
\end{equation}
where $\chi_0$ is a temperature-independent term, $C/T$ corresponds to the paramagnetic signal of defects (impurities) while the last term corresponds to a system of isolated dimers. $\Delta$ is the spin gap value and $N=2N_{\text{A}}$ since there are two vanadium atoms per formula unit. We found a good fit with $\chi_0=4.0(1)\cdot 10^{-4}$ emu/mol, $C=3.1(1)\cdot 10^{-3}$ emu$\cdot$K/mol, $g=1.55(2),\ \Delta=270.7(3)$~K. The $g$ value is somewhat lower than one can expect for V$^{+4}$ ($g=1.93-1.96$, see, for example, Ref.~\onlinecite{esr}). The underestimate of $g$ may be caused by the presence of the amorphous impurity in the samples under investigation. According to XRD (see above) the samples contain an appreciable ($\sim 25$\%) amount of the amorphous component that may really lead to an error in the sample weight and result in the decrease of $g$.\cite{foot2} The deviation of the spin system of \compound\ from the simple dimer model may be another reason for the decrease of $g$. However, the influence of this factor is much weaker than that of the amorphous component. We are convinced that a spin dimer model is the best choice for the reliable fitting of the present experimental data, since the use of the improved model (including four exchange integrals, see \ref{section-band}) makes the fit very unstable.

The subtraction of $\chi_0$ and of the paramagnetic contribution $C/T$ shows that the intrinsic susceptibility of \compound\ drops down to zero at low temperatures (see the inset of Fig.~\ref{suscept}) indicating a spin singlet ground state.

\subsection{Band structure and exchange interactions}
\label{section-band}
Fitting of experimental data provides a phenomenological way to study magnetic interactions. However, the crystal structure of \compound\ results in a rather complicated spin system with several different types of nearest-neighbor interactions (see Fig. \ref{structure}), therefore a mere phenomenological description of the system may appear unreliable. A microscopic picture will provide additional information about exchange interactions in the system.

\begin{figure}
\includegraphics{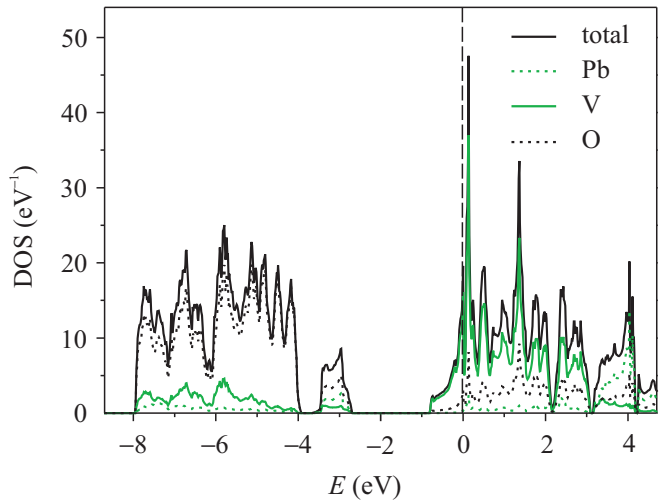}
\caption{\label{dos}(Color online) Total and atomic resolved density of states for the hypothetical PbV$_2$O$_5$ compound with the structure of \compound. The Fermi level is at zero energy.}
\end{figure}
A direct estimate of the exchange interactions can be derived on the basis of band structure data. The latter one is usually obtained by density-functional calculations. However, there is an additional difficulty in case of \compound\ since lead and cadmium atoms randomly occupy one crystallographic position. Sophisticated techniques (like CPA, VCA) or supercell calculations are required to treat correctly such type of disorder. Nevertheless, one may try a more simple way in order to study exchange interactions in \compound. According to the recent study of CaV$_2$O$_5$ and MgV$_2$O$_5$ \cite{korotin} the exchange integrals in these compounds are not sensitive to the type of metal cation (Ca or Mg) but depend on the geometry of the [V$_2$O$_5$] layer (namely, V--O--V angles). Therefore, we used the structural parameters of \compound\ (Table \ref{coordinates}) but set the occupancy factors $g_{\text{Pb}}=1,\ g_{\text{Cd}}=0$, or vice versa. Thus, the band structure for the two hypothetical compounds (PbV$_2$O$_5$ and CdV$_2$O$_5$) was calculated.

The density of states (DOS) plots for PbV$_2$O$_5$ and CdV$_2$O$_5$ are very similar, therefore we show only one of them in Fig.~\ref{dos}. The states below $-2$ eV are mainly formed by oxygen orbitals and provide the bonding between vanadium, lead or cadmium and oxygen. The highest occupied states have predominantly vanadium character with an admixture of oxygen. Note that gapless energy spectra of insulating transition metal compounds are a typical failure of LDA due to an underestimate of strong electron-electron correlations in the V $3d$ shell. The energy gap is readily reproduced by means of LSDA+$U$ (see, for example, Ref.~\onlinecite{korotin}).

\begin{figure}
\includegraphics{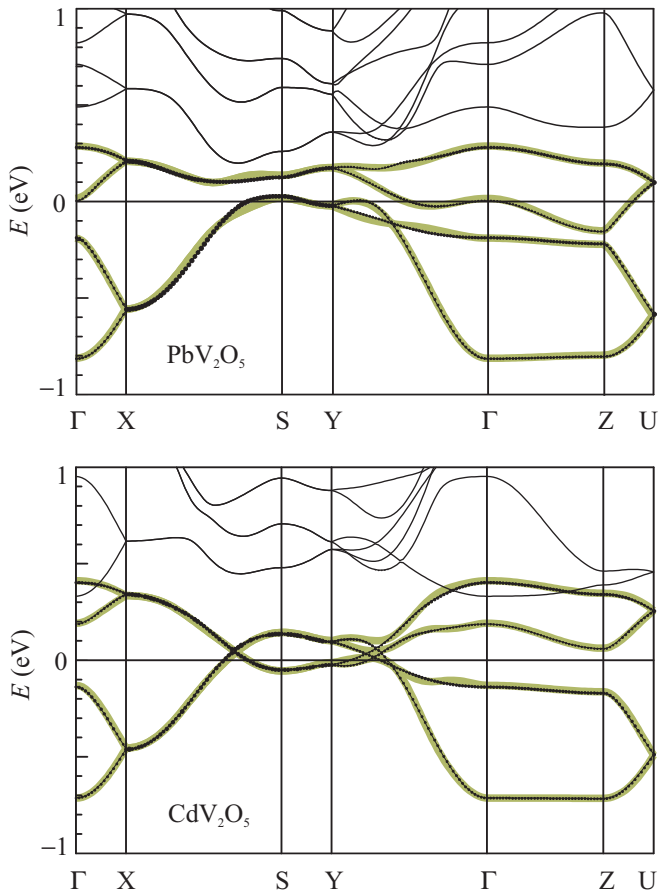}
\caption{\label{bands}(Color online) Band structure of PbV$_2$O$_5$ (upper panel) and CdV$_2$O$_5$ (lower panel) near the Fermi level. Black dots indicate the contribution of the V $3d_{xy}$ states. Thick solid lines show the fit of the tight-binding model to the LDA band structure.}
\end{figure}
According to simple crystal field considerations, the lowest (and hence occupied) vanadium $3d$ orbital in the V$^{+4}$O$_5$ square pyramid is the $d_{xy}$ one. There are four vanadium atoms in the unit cell of \compound, therefore we find near the Fermi level four bands formed by V $d_{xy}$ orbitals (Fig.~\ref{bands}). These bands are suitable for the tight-binding fit.

Similar four-band tight-binding models were constructed for both PbV$_2$O$_5$ and CdV$_2$O$_5$. The leading interactions of the models are shown in Fig.~\ref{structure}. A number of long-range interactions were also included in the model in order to provide a proper fit. Hopping parameters $t$ corresponding to the main interactions are listed in Table \ref{hopping}. Other $t$'s were found to be less than 0.020 eV, therefore one can neglect them while considering the overall magnetic behavior of the spin system. 

\begin{table}
\caption{\label{hopping}Tight-binding hopping parameters $t$ (in units of meV) and derived exchange integrals $J$ (in units of K)}
\begin{ruledtabular}
\begin{tabular}{cccccc}
(meV) & $t_1$ & $t_2$ & $t_3$ & $t_4$ & $t_{\perp}$ \\\hline
PbV$_2$O$_5$ & 159 & 63 & 72 & 57 & $-17$ \\
CdV$_2$O$_5$ & 147 & 26 & 52 & 98 & $-14$ \\
Averaged & 153 & 44.5 & 62 & 77.5 & $-15.5$ \\
\hline
(K) & $J_1$ & $J_2$ & $J_3$ & $J_4$ & $J_{\perp}$ \\
& 303 & 26 & 50 & 78 & 3 \\
\end{tabular}
\end{ruledtabular}
\end{table}
The hopping parameters of PbV$_2$O$_5$ and CdV$_2$O$_5$ are quite similar. If one considers the spin lattice in terms of coupled ladders, then the largest hopping ($t_1$) corresponds to the rungs of the ladders. Other nearest-neighbor hoppings are about half this value. The significant difference between the values of $t_2$ and $t_4$ for PbV$_2$O$_5$ and CdV$_2$O$_5$ may be caused by the influence of the metal cation on the corresponding interactions. Nevertheless, the basic feature of the magnetic interactions in both compounds is the same: strong coupling along the rungs of the ladder. Now, the $t$ values can be averaged and antiferromagnetic contributions to the exchange integrals are estimated as $J_i^{\afm}=4\bar t_i^2/U$ where $U$ is the effective on-site Coulomb repulsion. 

In general, one should also take into account ferromagnetic contributions since $J_i=J_i^{\afm}+J_i^{\fm}$. Unfortunately, the estimation of $J_i^{\fm}$ implies LSDA+U calculation, and the latter one is a difficult task for a partially disordered structure of \compound. According to Refs \onlinecite{korotin,korotin_prl} ferromagnetic interactions in layered vanadium oxides are relatively weak. For that reason we neglect $J_i^{\fm}$ and assume that $J_i=J_i^{\afm}$. Remarkable agreement between experimental and calculated values of $J_1$ (see below) \emph{a posteriori} justifies this approach.

We set $U=3.6$ eV according to Ref.~\onlinecite{korotin} and find $J_1^{\afm}=303$ K. This value may be compared with the spin gap $\Delta\approx 270$ K found by our fit of the susceptibility curve. In case of isolated dimers the spin gap is equal to the intradimer interaction and $J_1\approx\Delta\approx 270$ K in perfect agreement with the computational result. Other nearest-neighbor interactions are at least four times weaker than $J_1$, hence \compound\ may be considered as a system of weakly coupled dimers. The interlayer coupling is very weak ($J_{\perp}\approx 3$ K) as one can expect for the layered structure with the magnetically active $d_{xy}$ orbitals parallel to the layers.

\section{Discussion}
Experimental and computational data clearly show that \compound\ is similar to CaV$_2$O$_5$ and reveals a spin system of weakly coupled dimers. However, the intradimer interactions in these compounds differ by a factor of two. Now we will try to uncover a structural evidence of this change. The comparison of \compound\ with two known MV$_2$O$_5$ compounds (M = Ca, Mg) allows us to establish reliable correlations between exchange integrals and geometrical parameters.

The various magnetic interactions in the MV$_2$O$_5$-type compounds have different origin. $J_1$ and $J_2$ run via \mbox{V--O--V} superexchange paths, $J_3$ is a superposition of direct V--V exchange and $90^0$ \mbox{V--O--V} superexchange. $J_4$ corresponds to a more complicated superexchange path that involves two oxygen atoms and probably the M cation as well. Further we will focus on $J_1$ and $J_2$ since only single V--O--V paths may provide reasonably simple correlations between exchange integrals and geometrical parameters. In general, such paths are characterized by three parameters: V--O--V angle and two corresponding V--O distances. In MV$_2$O$_5$ compounds the parameters are related, therefore two parameters are sufficient: the angle and one distance (V--O or V--V). The geometrical parameters corresponding to $J_1$ and $J_2$ are listed in Table~\ref{exchange}.

\begin{figure}
\includegraphics{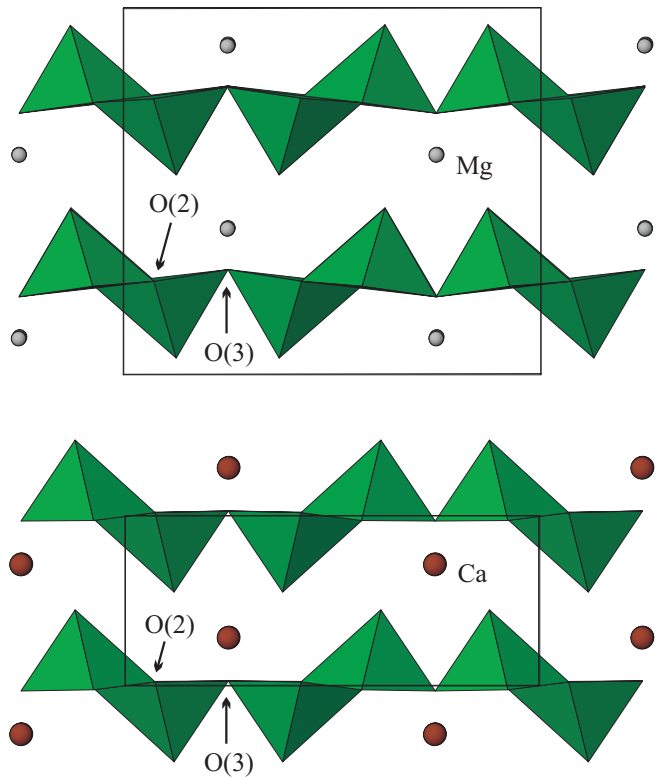}
\caption{\label{layers}(Color online) Comparison of the MgV$_2$O$_5$ (upper panel) and CaV$_2$O$_5$ (lower panel) structures. The small size of the Mg cation results in a corrugation of the [V$_2$O$_5$] layers.}
\end{figure}

\begin{table*}
\caption{\label{exchange}Comparison of nearest-neighbor exchange interactions and geometrical parameters for MV$_2$O$_5$ compounds}
\begin{ruledtabular}
\begin{tabular}{ccccccc}
& $J_1$ (K) & $d$(V--V) (\r A) & $\angle$VO(3)V (deg) & $J_2$ (K) & $d$(V--V) (\r A) & $\angle$VO(2)V (deg) \\
\compound & 303 & 3.550(5) & 132.1(3) & 26 & 3.667(7) & 137.3(2) \\
CaV$_2$O$_5$ & 608\footnotemark[1] & 3.493\footnotemark[2] & 132.9\footnotemark[2] & 122\footnotemark[1] & 3.604\footnotemark[2] & 135.3\footnotemark[2] \\
MgV$_2$O$_5$ & 92\footnotemark[1] & 3.372\footnotemark[3] & 117.6\footnotemark[3] & 144\footnotemark[1] & 3.692\footnotemark[3] & 141.1\footnotemark[3] \\
\end{tabular}
\end{ruledtabular}
\footnotetext[1]{Ref. \onlinecite{korotin_prl}}
\footnotetext[2]{Ref. \onlinecite{onoda}}
\footnotetext[3]{Ref. \onlinecite{onoda-mg}}
\end{table*}
The values in Table \ref{exchange} are easily understood if one considers the change of the crystal structure caused by the decrease of the cation size. In CaV$_2$O$_5$ calcium (ionic radius $r=1.26$ \r A) has eight-fold coordination and the [V$_2$O$_5$] layers are flat. Smaller cations like magnesium ($r=1.03$ \r A) require a decrease of the coordination number (six in case of Mg) that is achieved by the corrugation of the layers (see Fig.~\ref{layers}). Edge-sharing connections of the square pyramids are rigid in contrast to corner-sharing ones, therefore the corrugation of the layers results in a significant change of the V--O(3)--V angle. The averaged ionic radius of the effective \textquotedblleft Pb$_{0.55}$Cd$_{0.45}$\textquotedblright\ cation (1.33~\r A) is even larger than that of calcium and the layers remain flat. For that reason, angles remain almost constant but V--V distances (and hence V--O distances) increase. 

Now we turn to the exchange integrals. According to Ref.~\onlinecite{korotin} the dramatic decrease of $J_1$ in MgV$_2$O$_5$ compared to that in CaV$_2$O$_5$ is caused by the decrease of the \mbox{V--O(3)--V} angle. In case of \compound\ the decrease of $J_1$ is less pronounced since interatomic distances are changed instead of the angle. This result is quite reasonable as magnetic interactions are known to be far more sensitive to the variation of angles than to interatomic distances. 

The values of $J_2$ are more difficult to explain. The relevant bonding angles in the two compounds are quite similar. The V--V distance in MgV$_2$O$_5$ and \compound\ is higher than in CaV$_2$O$_5$ but $J_2\text{(MgV}_2\text{O}_5)\approx J_2\text{(CaV}_2\text{O}_5)\gg J_2$(\compound). One may suggest two reasons for such changes of $J_2$. First, tight-binding model reveals significant next-nearest-neighbor \textquotedblleft diagonal\textquotedblright\ interaction ($J_4\approx 80$ K) in \compound\ while lower estimates ($J_4\approx 20$ K) were given for CaV$_2$O$_5$ and MgV$_2$O$_5$.\cite{korotin,korotin_prl} This explanation seems to be reasonable (at least qualitatively) since lead and cadmium have different relevant orbitals compared to calcium or magnesium. Therefore, Pb and Cd may mediate superexchange interactions better. Note however that we discuss antiferromagnetic contributions to the exchange integrals only (see Band structure section), and ferromagnetic contributions may slightly change the situation. Second, the consideration of the geometrical parameters corresponding to V--O--V path only may be an oversimplification as other factors (e.g.~local environment of vanadium: all the V--O distances and O--V--O angles) are also important.

Thus, in case of complex superexchange paths it is difficult to impose a simple correlation between exchange integrals and geometrical parameters as the spin system in the MV$_2$O$_5$-type compounds is rather complicated. Nevertheless, we succeed in the explanation of the basic difference between \compound\ and CaV$_2$O$_5$. We found that the CaV$_2$O$_5$ structure is conserved rather well and the increase of the effective cation size in \compound\ results in a mere increase of interatomic distances. Both $J_1$ and $J_2$ are decreased and the type of the spin system (weakly coupled dimers) remains unchanged. 

The corrugation of the [V$_2$O$_5$] layer (alike the one that happens in MgV$_2$O$_5$) would be necessary to induce a really significant change of the spin system. The layers in CaV$_2$O$_5$ are almost flat and the further increase of the cation size can not give rise to their distortion. It means that one has to introduce a smaller (namely, smaller than calcium) cation between the [V$_2$O$_5$] layers in order to achieve an appreciable change of the spin system. The number of the appropriate cations is rather limited, therefore one may try to use a pair of different cations instead. The present study demonstrates that even two cations with fairly different size may randomly accommodate interstices between the vanadium-oxygen layers. This approach seems to be very promising for the search of new layered vanadium oxides since combinations of two different cations provide an easy way to adjust the cation size and modify the spin system in a controlled manner.

In conclusion, we prepared and investigated the complex vanadium oxide \compound. Its crystal structure is similar to that of CaV$_2$O$_5$. The new compound reveals a system of weakly antiferromagnetically coupled dimers with an intradimer interaction of about 270 K. In our approach using microscopic model with averaged hopping parameters we were able to describe the influence of structural changes on the magnetic interactions. Nearest-neighbor exchange interactions in the [V$_2$O$_5$] layers are decreased with the increase of the interatomic distances from CaV$_2$O$_5$ to \compound\ while the type of spin system remains unchanged. 
\begin{acknowledgments}
The authors acknowledge the financial support of RFBR (grant 07-03-00890), ICDD (GiA APS91-05), GIF (grant No. I-811-257.14/03) and the Emmy Noether Program. ZIH Dresden is acknowledged for computational facilities. A.Ts. is also grateful to the support of Alfred Toepfer Foundation and to MPI CPfS for hospitality.
\end{acknowledgments}


\begin{thebibliography}{27}
\bibitem[{foo({\natexlab{a}})}]{foot1}
\bibinfo{note}{Anisotropic broadening was observed for weak $(0k0)$
  reflections. It may be caused by slight inhomogeneities in the Pb/Cd
  distribution over the crystallites.}
\bibitem[{foo({\natexlab{b}})}]{foot2}
\bibinfo{note}{The weight error due to the presence of the amorphous phase in
  the samples under investigation may be revealed by fitting high-temperature
  susceptibility data with Curie-Weiss law and analyzing the value of the
  effective magnetic moment. We use the data in the temperature range of
  $350-400$ K and find reasonable fitting parameters ($\mu_{\text{eff}}=1.51\
  \mu_B,\ \Theta=-110$ K) that support our conclusions. The spin-only value of
  $\mu_{\text{eff}}$ for V$^{+4}$ is $1.73\ \mu_B$, while the experimental
  value is 13\% lower consistent with the decrease of $g$.}
\end{thebibliography}
\end{document}